# Deep learning based on PINN for solving 2 D0F vortex induced vibration of cylinder with high Reynolds number

Chen Cheng[1], Peng-Fei Xu[1], Yong-Zheng Li[3], Guang-Tao Zhang[2],

1. College of Harbor, Coastal and Offshore Engineering, Hohai University, Jiangsu, China
2. College of Mathematics and Informatics, South China Agricultural University, Guangdong, China
3. School of Naval Architecture and Ocean Engineering, Jiangsu University of Science and Technology, Jiangsu, China

## Abstract

Vortex-induced vibration (VIV) exists widely in natural and industrial fields. The main approaches for solving VIV problems are numerical simulations and experimental methods. However, experiment methods are difficult to obtain the whole flow field information and also high-cost while numerical simulation is extraordinary time-consuming and limited in low Reynolds number and simple geometric configuration. In addition, numerical simulations are difficult to handle the moving mesh technique. In this paper, physics informed neural network (PINN) is proposed to solve the VIV and wake-induced vibration (WIV) of cylinder with high Reynolds number. Compared to tradition data-driven neural network, the Reynolds Average Navier-Stokes (RANS) equation, by implanting an additional turbulent eddy viscosity, coupled with structure's dynamic motion equation are also embedded into the loss function. Training and validation data is obtained by computational fluid dynamic (CFD) technique. Three scenarios are proposed to validate the performance of PINN in solving VIV and WIV of cylinders. In the first place, the stiffness parameter and damping parameter are calculated via limited force data and displacement data; secondly, the flow field and lifting force/drag force are inferred by scattered velocity information; eventually, the displacement can be directly predicted only through lifting forces and drag forces based on LSTM. Results demonstrate that, compared with traditional neural network, PINN method is more effective in inferring and re-constructing the unknown parameters and flow field with high Reynolds number under VIV and WIV circumstances.

# 1. Introduction

Flow induced vibration (FIV) problems are ubiquitous in natural and industrial processes, such as pipeline in sea mining, cylinder of offshore and wind turbine etc. Vortex induced vibration (VIV) of bluff bodies, as a typical branch of FIV, will be happened when the vortex shedding frequency is close to the natural frequency of the structure. VIV can generate a huge amplitude vibration of structures according to the specific reduced velocity, the Reynolds number and structural dynamic characteristics (Blevins, 1990; Williamson and Govardhan, 2004). Sometimes, VIV can even cause large fatigue damage to the structures that attracts a substantial amount of attention.

Numerical simulation of VIV problems principally relies on solving the RANS equation and dynamic motion equation in a discretized form through finite element method (FEM), finite volume method (FVM) or finite difference method (FDM), which are described as computational fluid dynamic (CFD) method. However, CFD techniques are cumbersome in computational efficiency, especially for turbulent flow and complicated geometries. Furthermore, CFD techniques are also limitative in handling the moving mesh and other particular technical means.

Reduced order modeling (ROM), as one of the system identification, has been viewed as a strong tool to decrease the complexity and high dimensionality of the dynamical models and firstly proposed in optimal design, optimal control and inverse problem application. Proper orthogonal decomposition (POD) and dynamic mode decomposition (DMD) are two dominant methods of ROM in solving flow dynamics in lower dimensional representations (Dowell, 1997; Schmid, 2010). Henshawa et al (2007) utilized POD to construct the non-linear model of the aircraft behavior with low dimensionality and evaluate the performance on the real aircraft. Jovanovie et al (2014) developed a sparsity-promoting variant of the standard DMD algorithm to represent the flow field by numerical simulation and then compared to the experiments. The results showed that method can well re-construct the fluid model. Hemati et al (2014) formulated a low-storage approach to perform DMD to simulate the flow past cylinder and compared with the results from particle image velocimetry experiments. However, ROM also has limitations in solving complicated unsteady flows due to the information loss by compressive model. However, ROM makes fluid dynamics into the linear or weakly nonlinear problems with powerful assumptions which has limitation in complicated unsteady flow.

Deep learning (DL) technology has extraordinary ability to deal with the strong nonlinearity and high dimensionality (LeCun et al, 2015). Recently, DL has a tremendous breakthrough in some fields, such as speech recognition, image processing and event prediction (Goodfellow et al, 2016; Xiong et al, 2015). More recently, DL method is proposed for solving fluid dynamics. Ling et al (2016) constructed the deep learning of RANS turbulence model by embedding Galileo invariant into depth neural network, and firstly realized the prediction of channel flow vortex and separated flow. This is considered to be the first combination of deep neural networks and fluid mechanics (Nathan, 2017). Yeung et al (2017) proposed a deep learning framework for computing Koopman operators of nonlinear dynamic systems, which provides a new idea for modeling nonlinear dynamic systems by combining DMD method with deep neural networks. Miyanawala and Jaiman (2017) predicted the flow characteristics in the wake region of a two-dimensional cylinder by deep convolution network. Jin et al (2018) utilized fusion convolutional neural networks (CNNs) to predict the velocity fields around the circular cylinder by data obtained by pressure fields. Sekar et al (2019) also adopted CNNs technique combined with Multilayer Perceptron (MLP) to calculate the incompressible laminar steady flows. Recurrent neural network (RNN) is another powerful tool to predict temporal features of flow fields. Deng et al (2019) utilized the Long Short-Term Memory (LSTM) to obtain the time coefficient of the flow field. Mohan et al (2019) combined the CNNs and LSTM to predict the spatial-temporal features of turbulence dynamics. However, DL methods require magnanimous data to ensure the prediction accuracy and generalization ability. In addition, DL methods build up a surrogate model which is considered as black box and it means that the model lacks physical interpretation.

Raissi et al (2017) firstly proposed physics informed neural network (PINN) to solve the partial differential equations (PDE) and inverse problems. PINN modified the traditional form of the loss function and was embedded with the physical models, with its important breakthrough featuring that the PINN can predict the variables based on physical laws. Tartakovsky et al (2018) utilized PINN to construct the constitutive equations of Decay flow. It demonstrated that PINN has strong performance in solving inverse problems. Moreover, Yang et al (2020) employed Bayesian and PINN to solve the PDE with noisy data.

The aim of this paper is to utilize PINN method to solve the VIV and wake-induced vibration of cylinders. The turbulence eddy viscosity is introduced into the RANS model and then embedded into the loss function. A fully connected neural network and LSTM are adopted to construct the structure of the PINN. The whole flow field and unknown parameters (such as damping coefficient and stiffness coefficient) are calculated by PINN based on scattered training samples. The structure of paper can be demonstrated as follow. Section 2 introduces the governing equations of fluid mechanics and dynamic motion of the cylinders. Section 3 describes the principle of the FCNN and LSTM, then the scheme of PINN is introduced in Section 4. Section 5 demonstrate the three scenarios and show the performance of PINN in these scenarios. Conclusion is summarized in section 6.

## 2. Vortex induced vibration

*2.1 Governing equations of fluid mechanics*

The incompressible flow is governed by the Navier-Stokes (N-S) equation and conservation equation which can be shown as:

$$\mathcal{F}(\mathbf{u}, p) = 0 \Rightarrow \begin{cases} \nabla \cdot \mathbf{u} = 0 & x, t \in \Omega_{f,t}, \theta \in \mathbb{R}^d \\ \dfrac{\partial \mathbf{u}}{\partial t} + (\mathbf{u} \cdot \nabla)\mathbf{u} + \dfrac{1}{\rho}\nabla p - \nu \nabla^2 \mathbf{u} + \mathbf{b}_f = 0, & x, t \in \Omega_{f,t}, \theta \in \mathbb{R}^d \end{cases} \quad (1)$$

where **u** denotes the velocity field (including $u$, $v$, $w$); $p$ the pressure field; $v$ the kinematic viscosity; $\mathbf{b}_f$ the body force.

Incompressible flow can be solved with the proper initial and boundary conditions by the numerical simulation. However, the turbulent flows are generated with the increasing Reynolds number, the N-S equation is difficult to be solved directly owing to not only huge computational expense, but also the illness or stiffness of the algebraic matrices involved (Durbin, 2018). Reynolds-average N-S (RANS), as a strong tool in industrial practices, is proposed to solve the turbulent flow. The governing equation can be demonstrated as follow:

$$\frac{\partial \bar{u}_i}{\partial x_i} = 0$$

$$\frac{\partial \rho_f \bar{u}_i}{\partial t} + \frac{\partial \rho_f \overline{u_i u_j}}{\partial x_i} = -\frac{\partial \bar{p}}{\partial x_i} + \mu \nabla^2 \bar{u}_i - \frac{\partial \rho_f \overline{u'_i u'_j}}{\partial x_j} \quad (2)$$

where:

$$-\rho \overline{u'_i u'_j} = \mu_t \left( \frac{\partial u_i}{\partial x_j} + \frac{\partial u_j}{\partial x_i} \right) - \frac{2}{3} \rho k_t \delta_{ij} \quad (3)$$

where $-\rho \overline{u'_i u'_j}$ denotes Reynold stress $\tau_{ij}$; $\overline{(\cdot)}$ the Reynolds average or the spatial filtering, and $u'_i = u_i - \bar{u}_i$. According to the Fick's law, Reynold stress can be re-modelled as:

$$\rho \overline{u'_i u'_j} = \rho v_t \frac{\partial \bar{u}_i}{\partial x_j} \quad (4)$$

where $v_t$ denotes the turbulent eddy viscosity. The value of $v_t$ is determined by the flow filed. The parameter $v_t$ has been calibrated by various methods for several decades (Poroseva et al, 2016). It is difficult to obtain the $v_t$ in a universal sense due to the case-by-case dependence. Fortunately, a great amount of practices shows that modelling eddy viscosity can well establish the fitting between filtered experimental data and solutions of the RANS. With this ideal, the traditional RANS equation can be re-modeled as follow (Bai et al, 2021):

$$\frac{\partial \bar{u}}{\partial t} + \bar{u} \frac{\partial \bar{u}}{\partial x} + \bar{v} \frac{\partial \bar{u}}{\partial y} = -\frac{\partial \bar{p}}{\partial x} + (v + v_t) \left( \frac{\partial^2 \bar{u}}{\partial x^2} + \frac{\partial^2 \bar{u}}{\partial y^2} \right)$$

$$\frac{\partial \bar{v}}{\partial t} + \bar{u} \frac{\partial \bar{v}}{\partial x} + \bar{v} \frac{\partial \bar{v}}{\partial y} = -\frac{\partial \bar{p}}{\partial y} + (v + v_t) \left( \frac{\partial^2 \bar{v}}{\partial x^2} + \frac{\partial^2 \bar{v}}{\partial y^2} \right) \quad (5)$$

$$\frac{\partial \bar{u}}{\partial x} + \frac{\partial \bar{v}}{\partial y} = 0$$

*2.2 Kinematic equation and discrete method*

The oscillation of the cylinder can be described as a typical mass-spring-damper system. Therefore, the motion equations of cylinder in x-direction and y-direction can be shown as follow:

$$m_{system}\frac{d^2\eta}{dt^2}+c\frac{d\eta}{dt}+k\eta=F_L(t)$$
$$m_{system}\frac{d^2\varsigma}{dt^2}+c\frac{d\varsigma}{dt}+k\varsigma=F_D(t) \quad (6)$$

where $m_{system}$ denotes the inertial mass of the vibration system; $\varsigma$ and $\eta$ the motion in x-direction and y-direction, respectively; $c$ the damping and $k$ the stiffness coefficient; $F_L(t)$ and $F_D(t)$ represent the lift force and drag force, respectively.

The velocities of the oscillatory cylinder can be calculated as follow:

$$v_x(t)=\dot\varsigma(t)$$
$$v_y(t)=\dot\eta(t) \quad (7)$$

The initial conditions and boundary conditions of the oscillated cylinder can be described as:

$$v_x(0)=v_y(0)=0$$
$$x(0)=y(0)=0 \quad (8)$$

When the lift force and drag force are calculated by integrating the pressure and velocity gradients, the Eq. 6 can be discretized by the fourth-order Runge-Kutta method. The equation in y-direction can be expressed as follow:

$$\dot\eta(t_{n+1})=\dot\eta(t_n)+\frac{\Delta t}{6}(k_1+2k_2+2k_3+k_4)$$
$$\eta(t_{n+1})=\eta(t_n)+v_y(t_n)\Delta t+\frac{\Delta t^2}{6}(k_1+k_2+k_3) \quad (9)$$

where:

$$k_1=\frac{F_L(t_n)}{m_{system}}-\frac{c}{m_{system}}v_y(t_n)\Delta t-\frac{k}{m_{system}}\eta(t_n)$$
$$k_2=\frac{F_L(t_n)}{m_{system}}-\frac{c}{m_{system}}\left[v_y(t_n)+\frac{\Delta t}{2}k_1\right]-\frac{k}{m_{system}}\left[\eta(t_n)+\frac{\Delta t}{2}v_y(t_n)\right]$$
$$k_3=\frac{F_L(t_n)}{m_{system}}-\frac{c}{m_{system}}\left[v_y(t_n)+\frac{\Delta t}{2}k_2\right]-\frac{k}{m_{system}}\left[\eta(t_n)+\frac{\Delta t}{2}v_y(t_n)+\frac{\Delta t^2}{4}k_1\right]$$
$$k_4=\frac{F_L(t_n)}{m_{system}}-\frac{c}{m_{system}}\left[v_y(t_n)+k_3\Delta t\right]-\frac{k}{m_{system}}\left[\eta(t_n)+\frac{\Delta t}{2}v_y(t_n)+\frac{\Delta t^2}{2}k_2\right] \quad (10)$$

where $k_1$、$k_2$、$k_3$ and $k_4$ are the coefficients of the fourth-order Runge-Kutta, $\Delta t$ denotes the time step; $v_y$ the velocity of the cylinder in y-direction. The equation in

*x*-direction is same as that in *y*-direction.

## 3. Deep learning

*3.1 Fully connected neural network*

Fully connected neural network (FCNN) is a classic structure of neural network. FCNN consist input layer, hidden layer and output layer which can be viewed in Fig. 1. Layer 0 is input layer and layer L is output layer, the other layers are hidden layers. Each layer includes large number of neurons which have weights, biases and activation functions. The weights and biases are tuned by training the neural network. It is noteworthy that the activation function plays a significant role in handling the nonlinear problems. The common activation functions are sigmoids, tanh and rectified linear units. In the recent year, the adaptive activation function is demonstrated has better performance in solving strong nonlinearity (Jagtab et al, 2019). The output of a neuron can be calculated as follow:

$$z_j^l = w_{jk}^l f_{l-1}\left(z_k^{l-1}\right) + b_j^l \tag{11}$$

where $z_j^l$ denotes the output of neuron *j* in layer *l*; $w_{jk}^l$ the weight between neuron *k* in layer *l*-1 and neuron *j* in layer *l*; $f(\cdot)$ the activation function; $b_j^l$ the bias of neuron *j* in layer *l*. The formula can also be written as:

$$z_j^l = \sum_k w_{jk}^l y_k^{l-1} + b_j^l \tag{12}$$

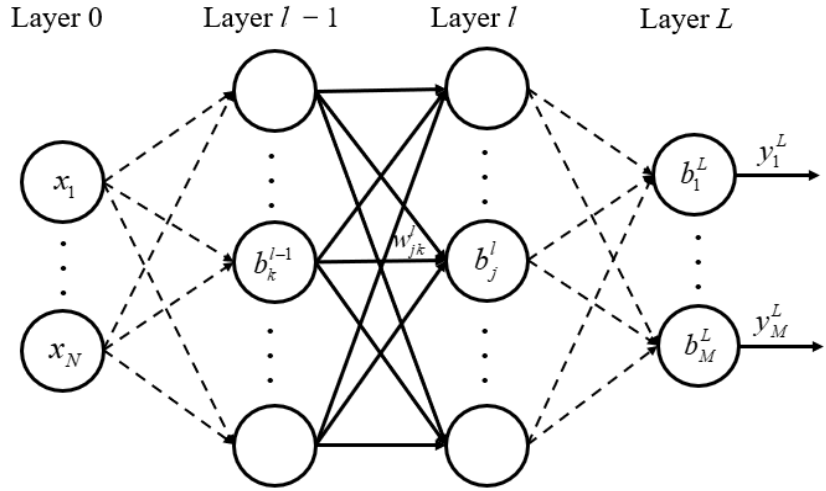

Fig. 1. The structure of the fully-connected neural network

The input data $x = [x_1, x_2, \cdots x_n]$ and output data $y = [y_1, y_2, \cdots y_n]$ are utilized to adjust the parameters in neural structure including weights and biases. The approximate result $y^L$ predicted by FCNN is compared to the exact value $y$, the difference between approximate result and exact result is defined as cost function which can be

shown as:

$$w^*, b^* = \arg\min_{w,b} C(y, y^L) \tag{13}$$

where $w^*$ and $b^*$ are tuned weights and biases, respectively; $C(y, y^L)$ represents the cost function. How to reduce the loss function as much as possible is the premise to ensure that the neural network can effectively predict the concerned results. Generally, backpropagation is a standard approach to compute the gradients and can be viewed as follow:

$$\delta_j^l = \frac{\partial C}{\partial z_j^l} \tag{14}$$

Go a step further, the gradient of the cost function can be computed as another form which can be demonstrated:

$$\begin{aligned} \delta_j^L &= \frac{\partial C}{\partial y_j^L} \sigma_L'(z_j^L), & \frac{\partial C}{\partial w_{jk}^l} &= y_k^{l-1} \delta_j^l, \\ \delta_j^l &= \sum_k w_{kj}^{l+1} \delta_k^{l+1} \sigma_l'(z_j^L), & \frac{\partial C}{\partial b_j^l} &= \delta_j^l \end{aligned} \tag{15}$$

The $\delta$-term in Eq. (5) can be expressed as vector form:

$$\delta^L = \nabla_{y^L} C \odot \sigma_L'(z^L), \quad \delta^l = (W^{l+1})^T \delta^{l+1} \odot \sigma_l'(z^l) \tag{16}$$

$$\nabla_{y^L} C = \left[ \frac{\partial C}{\partial y_1^L}, \cdots, \frac{\partial C}{\partial y_M^L} \right]^T \tag{17}$$

where $\odot$ is the Hadamard product. The notation $\nabla C$ without a subscript the vector of partial derivatives in respect of the input $x = [x_1, x_2, \cdots x_n]$.

*3.2 Long short-term memory*

Long short-term memory (LSTM) is a kind of time series neural network, which is specially designed to solve the long-term dependence problem of general RNN (recurrent neural network). The advantage of LSTM is to store and memorize previous information which can reduce the complexity and number of layers in its structure (Wang, 2017).

The structure of LSTM includes input gate, forget gate, block input, cell state, output gate and block output. By controlling to open and close the gates, LSTM is able to truncate gradients in the neural network. The model of the LSTM can be described as follow:

$$\begin{aligned}
i_t &= \sigma\left(W_{x_i} x_t + W_{h_i} h_{t-1} + b_{i_i}\right) \\
f_t &= \sigma\left(W_{x_f} x_t + W_{h_f} h_{t-1} + b_{i_f}\right) \\
z_t &= \tanh\left(W_{x_c} x_t + b_c\right) \\
c_t &= f_t \odot c_{t-1} + i_t \odot z_t \\
o_t &= \sigma\left(W_{x_o} x_t + W_{h_o} h_{t-1} + b_{i_o}\right) \\
h_t &= o_t \odot \tanh(c_t)
\end{aligned} \qquad (18)$$

where $x_t$ denotes the input vector at time step $t$; $W_{(\cdot)}$ and $b_{(\cdot)}$ the weight matrix and threshold vector respectively; $\sigma(\cdot)$ the activation function; $\odot$ is the Hadamard product; LSTM can control the flow of data information via opening and closing the different gates which can be demonstrated in Fig. 2.

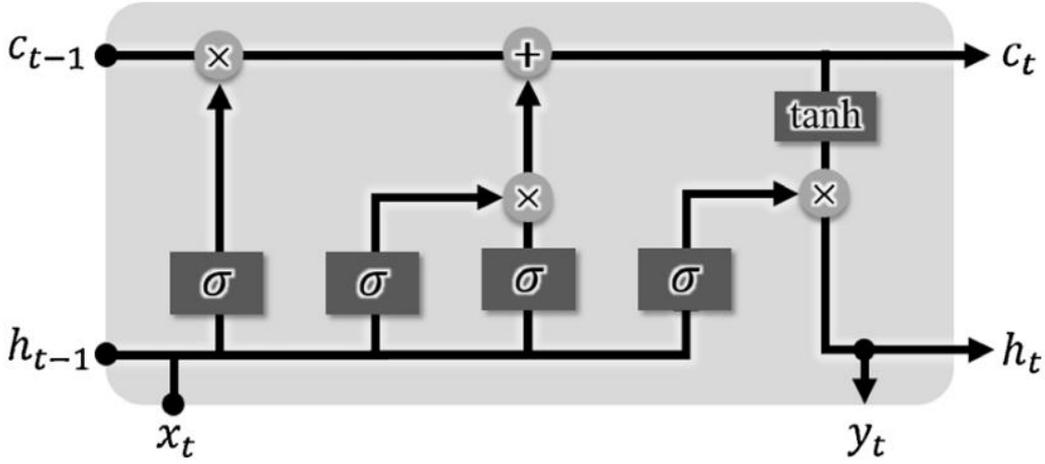

Fig. 2. The structure of the long short-term memory

## 4. Physics informed neural network

### 4.1 Physics-constrained deep learning

Conventionally, DL method builds up a surrogate model, such as FCNN or CNN, for predicting the solution of the fluid flow which are approximately equal to real values (Zhu et al, 2018).

$$\mathbf{f}(t, x, \theta) \approx \tilde{\mathbf{f}}(t, x, \theta) \triangleq \mathbf{z}_l(t, x, \theta; \mathbf{W}, \mathbf{b}) \qquad (19)$$

where $\mathbf{f}$ is the solution vector including the velocity fields and pressure fields; $\mathbf{W}$ and $\mathbf{b}$ denote the weights and biases, respectively. $\mathbf{z}_l(t, x, \theta; \mathbf{W}, \mathbf{b})$ the predicted by the surrogate model; $\tilde{\mathbf{f}}$ the locally minimized. The solution of flow dynamics can be cast into an optimization problem which can be demonstrated as follow:

$$\mathcal{L}_{data}(\mathbf{W},\mathbf{b}) = \frac{1}{N}\left|\mathbf{f}^d(t,\mathbf{x},\theta) - \mathbf{z}_l(t,\mathbf{x},\theta;\mathbf{W},\mathbf{b})\right|^2$$
$$\mathbf{W}^*, \mathbf{b}^* = \underset{\mathbf{w},\mathbf{b}}{argmin}\,\mathcal{L}_{data}(\mathbf{W},\mathbf{b}) \quad (20)$$

where $\mathcal{L}_{data}(\mathbf{W},\mathbf{b})$ denotes the loss function based on data; N the number of training samples. $\mathbf{f}^d$ the training data.

However, the traditional DL requires large number of training data, which is too difficult to achieve from time-consuming CFD simulation. Physics-constrained deep learning embeds the physical model into the loss function by minimizing the violation of the solution on the basis of the known partial differential equations for fluid flows over a domain of interests without the demands of handling these equations for each parameter with conventional numerical simulations. The residual of N-S equations and mass conservation equations are computed by FCNN and the specific loss function can be demonstrated as follow:

$$\mathcal{L}_{phy}(\mathbf{W},\mathbf{b}) = +\underbrace{\frac{1}{N_f}\sum_{i=1}^{N_f}\left|\frac{\partial \mathbf{u}}{\partial t} + (\mathbf{u}\cdot\nabla)\mathbf{u} + \frac{1}{\rho}\nabla p - \nu\nabla^2\mathbf{u} + \mathbf{b}_f\right|^2}_{Structure\ imposed\ by\ N-S\ equations} + \underbrace{\frac{1}{N_f}\sum_{i=1}^{N_f}\left|\nabla\mathbf{u}\right|^2}_{Mass\ conservation}$$

$$\mathbf{W}^*, \mathbf{b}^* = \underset{\mathbf{w},\mathbf{b}}{argmin}\,\mathcal{L}_{phy}(\mathbf{W},\mathbf{b}) \quad (21)$$

$$s.t.\begin{cases}\mathcal{I}(\mathbf{x},p,\mathbf{u},\theta) = 0, & t=0,\ in\ \Omega_f \\ \mathcal{B}(t,\mathbf{x},p,\mathbf{u},\theta) = 0 & on\ \partial\Omega_{f,t}\end{cases}$$

where $\mathcal{L}_{phy}(\mathbf{W},\mathbf{b})$ denotes the physics-based loss; $\mathcal{I}$ and $\mathcal{B}$ the initial and boundary conditions, respectively;

The first and/or second derivative terms of velocity and pressure in the loss function can be computed by the automatic differentiation approach (AD) (Baydin et al, 2018). Compared to the traditional differential calculation, such as Manual Differentiation, Numerical Differentiation and Symbolic Differentiation, the core problem of AD is to calculate the derivatives, gradients and Hessian matrix values of complex functions, which are usually multi-layer composite functions at a certain point. The advantage of the AD is more accurate due to the absence of truncation or round-off errors. Generally, AD can be directly utilized in deep learning framework such as Tensorflow, Pytorch and Theano (Paszke et al, 2017; Abadi et al, 2016; Bastien et al, 2012). In order to reduce the error of the loss function, the Adam optimizer is utilized to optimize the target function. Adam optimizer can constantly adjust the learning rates with the situation changes in the learning process (Diederik and Jimmy, 2017). 'Xavier' method is designed to decide the initial weights and biases which can ensure faster convergence of neural network (Glorot and Bengio, 2010). A residual neural network is added in the FCNN to avoid gradient explosion and/or gradient disappearance (He et al, 2016).

*4.2 Initial and Boundary condition enforcement*

The loss function constrained by the physical equations becomes identically zero, the predicted values of velocity and pressure fields will precisely satisfy the N-S equations. Consequently, the solutions driven by FCNN particularly have physical interpretation through penalizing the PDE residuals. Furthermore, to make the problem well-posed, the appropriate initial conditions and boundary conditions are required and imposed as constraints which are dealt with a soft manner by amending the original loss function with penalty terms (Márquez-Neila et al, 2017). The Eq. (10) can be rewritten by adding initial loss and boundary loss as follow:

$$\mathcal{L}_{phy}^{c}(\mathbf{W},\mathbf{b},\lambda_i,\lambda_b) = \underbrace{\mathcal{L}_{phy}(\mathbf{W},\mathbf{b})}_{Equation\ loss} + \underbrace{\lambda_i \|\mathcal{I}(\mathbf{x},p,\mathbf{u},\theta)\|_{\Omega_{f,t}}}_{Initial\ loss} + \underbrace{\lambda_b \|\mathcal{B}(t,\mathbf{x},p,\mathbf{u},\theta)\|_{\partial\Omega_{f,t}}}_{Boundary\ loss} \quad (22)$$

where $\lambda_i$ and $\lambda_b$ are penalty coefficients.

## 5. PINN for solving vortex induced vibration of 2DOF cylinder

*5.1 CFD method for obtaining data*

CFD techniques of the 2 DOF of VIV and WIV are carried out and the simulation results are selected as training data. It is interesting to note that the experimental data can also be utilized for training neural network. 2D flow field is calculated through the solver, pimpleDyMFoam, executed in OpenFOAM, which is an open source framework of FVM. Shear stress transport (SST) $k-\omega$, as a known turbulence model is employed. Furthermore, the nested grid technique, which is the latest dynamic grid, is adopted to handle the moving boundary of cylinder.

The whole computational zone is a rectangle region with the length of 40$D$ and the width of 20$D$. $D$ is the diameter of cylinder and located in origin of coordinate. The inlet flow is enforced on the left part of the computational zone with a Dirichlet boundary condition $u=(U_\infty,0)$ while the outlet is Neumann boundary (zero-gradient pressure) at the right part of computational zone. The distance between the inlet and the center of the cylinder is 10$D$ while the distance between the outlet and the center of the cylinder is 30$D$, that can guarantee that the cylinder is not distributed by remoting boundary. The upper and bottom part of computational zone is slide-wall. The Reynolds number, $Re = U_\infty D/\nu$, is 2889. The parameters of cylinder are the damping parameter $c = 0.07444$ and the stiffness parameter $k = 17.2589$. The concrete details can be viewed in Fig. 3.

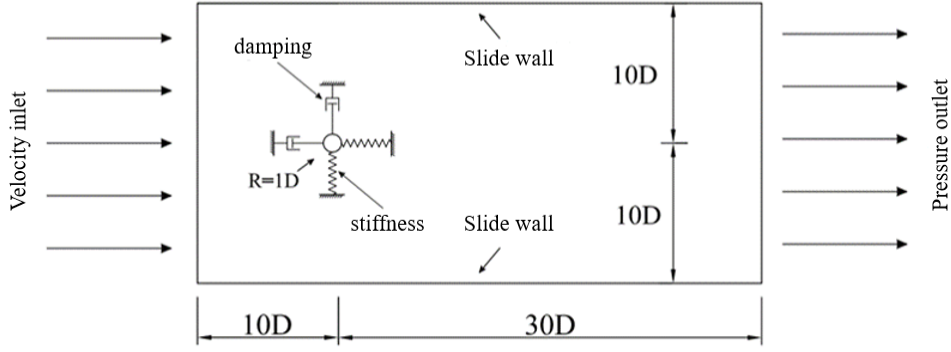

Fig. 3. The sketch of the vortex induced vibration of cylinder

*5.2 Inferring damping and stiffness parameters from forces and displacement*

It is fact that the stiffness and damping parameters of cylinder cannot be measured directly but the force and displacement of cylinder can be measured by force balance and laser range finder. Therefore, the PINN method is adopted to infer the damping and stiffness parameter through the limited force and displacement datasets. The PINN for solving parameters of cylinder can be viewed in Fig. 4 and the physical law can be described as follow:

$$F_L := m\eta_{tt} + c\eta_t + k\eta$$
$$F_D := m\varsigma_{tt} + c\varsigma_t + k\varsigma \tag{23}$$

It is noteworthy that the damping and stiffness parameters are transformed into the parameter of the resulting PINN. The loss function can be viewed as follow:

$$Loss = \|\eta(t^n) - \eta^n\| + \|\varsigma(t^n) - \varsigma^n\| + \|F_L(t^n) - F_L^n\| + \|F_D(t^n) - F_D^n\| \tag{24}$$

Training dataset ($N=120$), corresponding to exact damping and stiffness parameters, is utilized to tune a neural network with 10 hidden layers and 32 neurons in each layer by minimizing the sum of mean squared error of loss function utilizing the Adam and L-BGFS optimizer. The PINN is applied to predict the whole solution functions of forces and displacements in two degrees, as well as the stiffness and damping parameters. The predicted values for stiffness and damping parameter are $k = 17.1983$ and $c = 0.07138$. This corresponds to about 0.62% and 0.02% comparative errors in the predicted results for *k* and *c*, respectively. The entire functions of forces and displacement can also well be inferred by PINN which can be viewed in Fig. 5. It is obvious that the PINN can obtain the unknown parameters of cylinder effectively from limited dataset and also achieve the entire force and displacement functions.

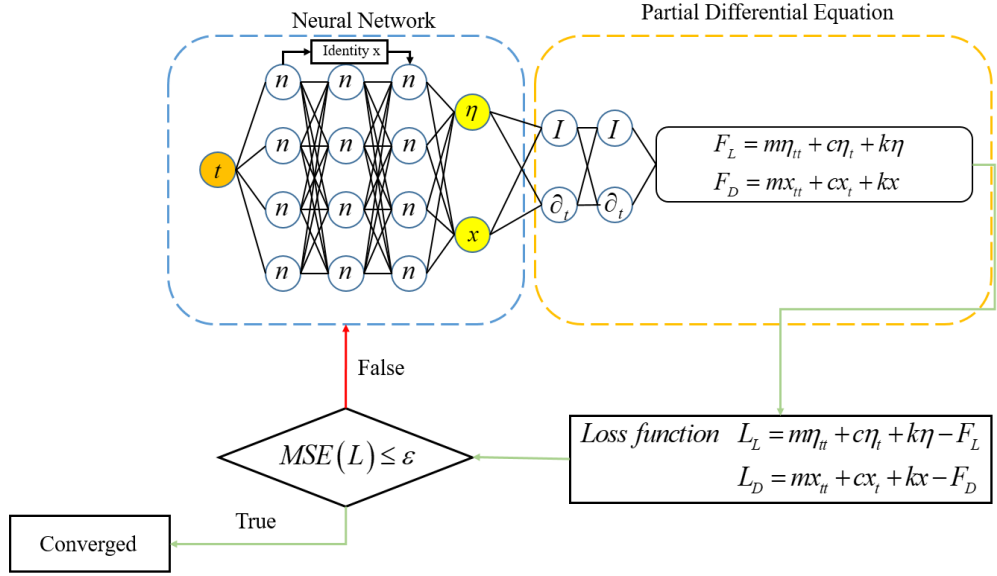

Fig. 4. PINN method for solving damping and stiffness parameters

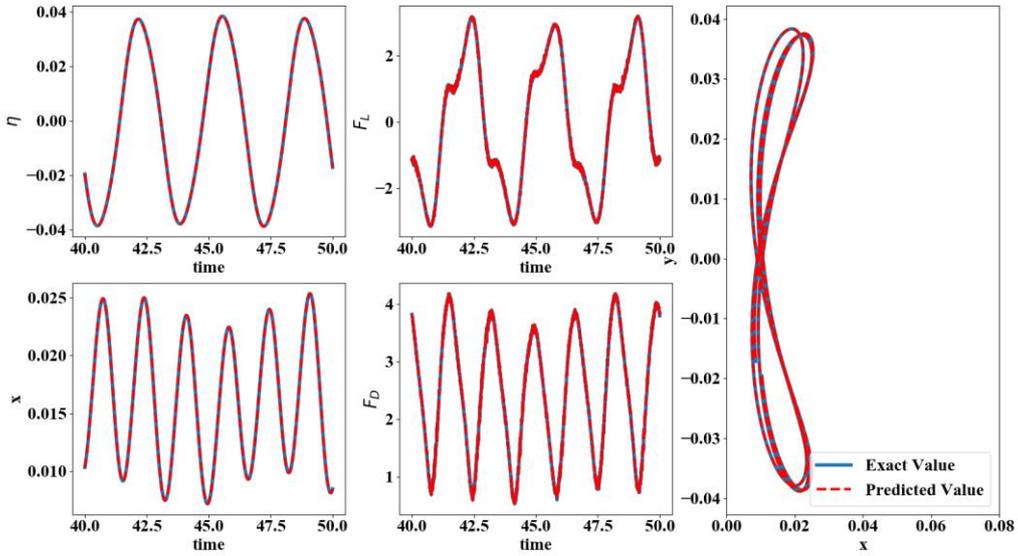

Fig. 5. PINN method for predicting the force and displacement

*5.3 Inferring lift force, drag force and pressure from scatter velocity field*

The aim of this section is to reconstruct the whole flow field with high Reynolds number and infer the forces (including lifting force and drag force) enforcing on the moving cylinder based on scattered information $\{t^n, \eta^n, \varsigma^n\}$ and $\{t^n, x^n, y^n, u^n, v^n\}$. It is worth recalling that the pressure information and turbulent eddy viscosity are viewed unknown parameters that also need to be solved. The loss function induced by partial differential equations includes three parts which are contributed by the *x*-component velocity $u$, *y*-component velocity $v$ and mass conversation, respectively. It can be demonstrated as follow:

$$e_1 = \frac{\partial u}{\partial t} + u\frac{\partial u}{\partial x} + v\frac{\partial u}{\partial y} + \frac{\partial p}{\partial x} - (v+v_t)\left(\frac{\partial^2 u}{\partial x^2} + \frac{\partial^2 u}{\partial y^2}\right) + \varsigma_{tt}$$

$$e_2 = \frac{\partial v}{\partial t} + u\frac{\partial v}{\partial x} + v\frac{\partial v}{\partial y} + \frac{\partial p}{\partial y} - (v+v_t)\left(\frac{\partial^2 v}{\partial x^2} + \frac{\partial^2 v}{\partial y^2}\right) + \eta_{tt} \qquad (25)$$

$$e_3 = \frac{\partial u}{\partial x} + \frac{\partial v}{\partial y}$$

For the simplicity of presentation, the over-line symbol for the operator in Eq.2 is omitted. It should be noted that the horizontal displacement and vertical displacement are incorporated into $e_1$ and $e_2$, respectively, so that the fluid flow coordinate system is attached to the cylinder. The total loss function can be determined as:

$$\begin{aligned}L_{sum} &= L_u + L_v + L_\eta + L_\varsigma + L_e \\ &= \|u(t^n, x^n, y^n) - u^n\| + \|v(t^n, x^n, y^n) - v^n\| \\ &+ \|\eta(t^n) - \eta^n\| + \|\varsigma(t^n) - \varsigma^n\| + \sum_{i=1}^{3}\|e_i\|\end{aligned} \qquad (26)$$

The specific process of PINN to solve this problem can be viewed in Fig. 6. The fully-connected neural network including 12 hidden layers with 32 neurons in each layer computes the gradient of loss function via Adam optimizer. The differential operations are generated by automatic differentiation implemented in Tensorflow. The adaptive activation function is adopted in each layer to enhance the nonlinear processing capability. The Latin hypercube sampling (LHS) is adopted to obtain the training data in different snapshots, the total numbers of training data are 40000. Three snapshots of flow field ($t = 30s, 40s, 50s$) are selected to validate the PINN

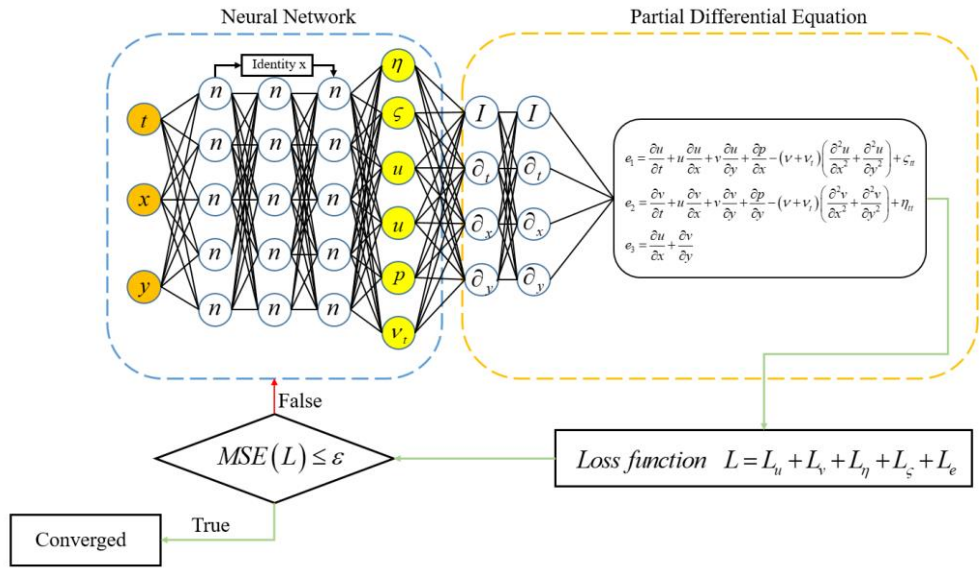

Fig. 6. PINN method for solving lift force, drag force and entire fluid flow

Fig. 7 demonstrates the whole flow field simulated by CFD technique and inferred flow field by PINN at different snapshots. It is obvious that the proposed framework is able to reconstruct the whole velocity field with high Reynold number accurately which can be viewed in Fig. 7(*a*) and Fig. 7(*b*). A remarkable result originates from PINN's ability to infer the whole pressure field accurately in defect of any training samples on the pressure itself (view Fig. 7(*c*)). The mean square errors of velocity field and pressure field are listed in Table 1. A strange phenomenon is that the difference in magnitude between the predicted pressure and exact one, although the distribution of the pressure filed is almost same. It is validated by the law of the N-S equation due to the pressure field is only recognizable up to a fixed value. For the incompressible flow, the absolute value of the pressure is of no great important.

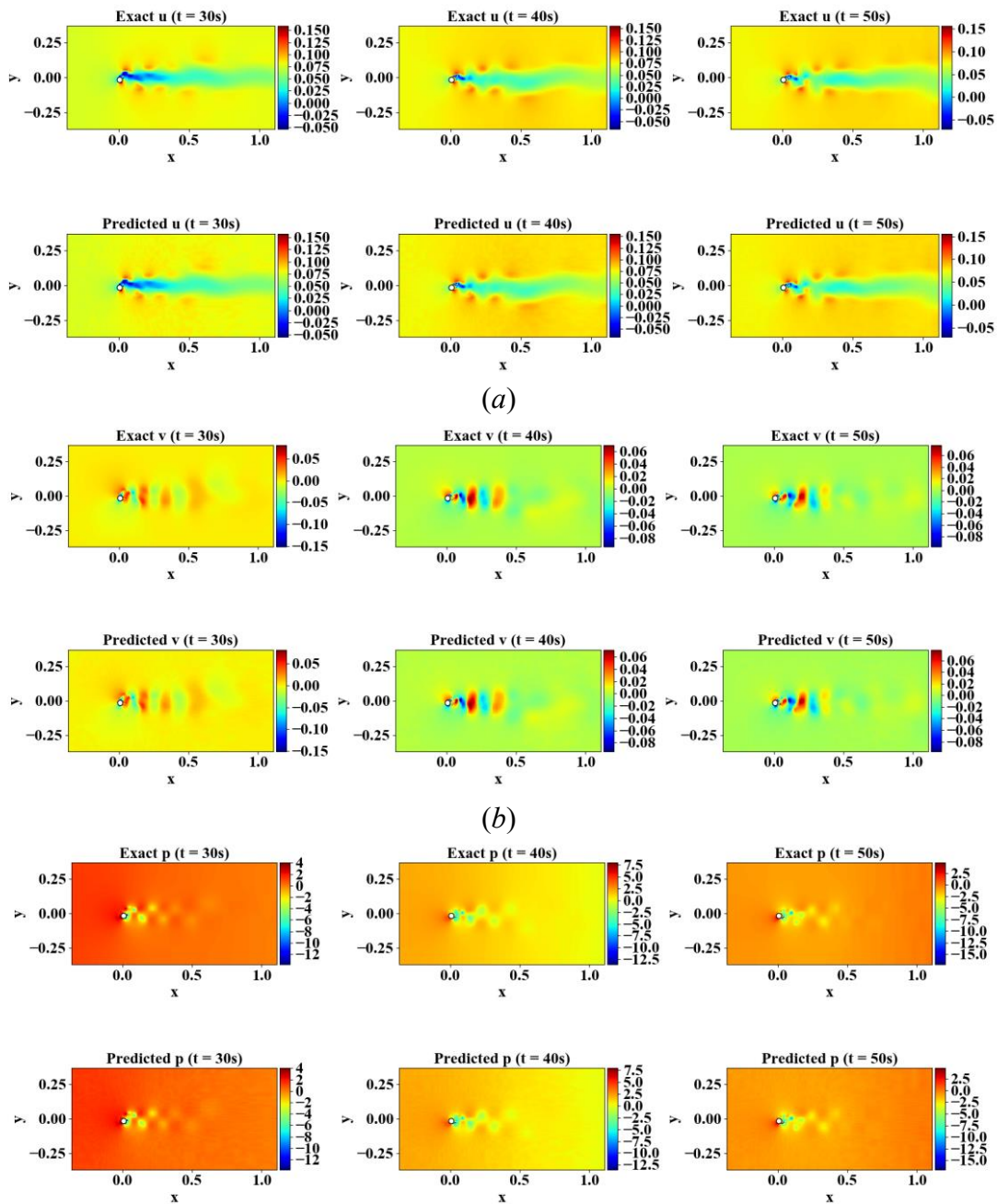

(*a*)

(*b*)

(c)

Fig. 7. PINN method for reconstructing the flow field with different times ((a) u, (b) v, (c) p)

Table 1. Mean square errors of entire flow fields

| | Mean square errors | | |
|---|---|---|---|
| | $u$ | $v$ | $p$ |
| $t = 30s$ | $2.83 \times 10^{-3}$ | $0.82 \times 10^{-3}$ | $0.4 \times 10^{0}$ |
| $t = 40s$ | $1.62 \times 10^{-3}$ | $1.36 \times 10^{-3}$ | $0.87 \times 10^{0}$ |
| $t = 50s$ | $1.42 \times 10^{-3}$ | $0.92 \times 10^{-3}$ | $0.21 \times 10^{0}$ |

When the pressure and velocity fields are obtained, the lift force and drag force on the cylinder can be approximately calculated, based on the function of the pressure and velocity gradients, through trapezoidal law as:

$$\begin{aligned} F_L &= \oint_\Gamma \left[ -pn_y + \frac{2}{Re}\frac{\partial v}{\partial y}n_y + \frac{1}{Re}\left(\frac{\partial u}{\partial y} + \frac{\partial v}{\partial x}\right)n_x \right] ds \\ F_D &= \oint_\Gamma \left[ -pn_x + \frac{2}{Re}\frac{\partial u}{\partial x}n_x + \frac{1}{Re}\left(\frac{\partial u}{\partial y} + \frac{\partial v}{\partial x}\right)n_y \right] ds \end{aligned} \quad (27)$$

where $(n_x, n_y)$ denotes the outward normal on the cylinder while $ds$ the arc length on the surface of the cylinder.

Fig. 8 shows the comparison between the inferred drag and lifting forces with the exact ones. The blue solid lines represent exact values while the orange dotted line represent inferred values. The mean square of errors of lifting force and drag force are $0.17 \times 10^{-4}$ and $0.63 \times 10^{-4}$. PINN can well calculate the forces on cylinder due to the accurate prediction of velocity and pressure fields and then infer the damp and lifting coefficients by above setup. Therefore, in practical engineering, we only utilize the particle image velocimetry (PIV) to obtain the scattered velocity information that can infer the whole flow field and forces on structures. It is no doubt that it greatly reduces the difficulty of obtaining the experimental data of vortex induced vibration.

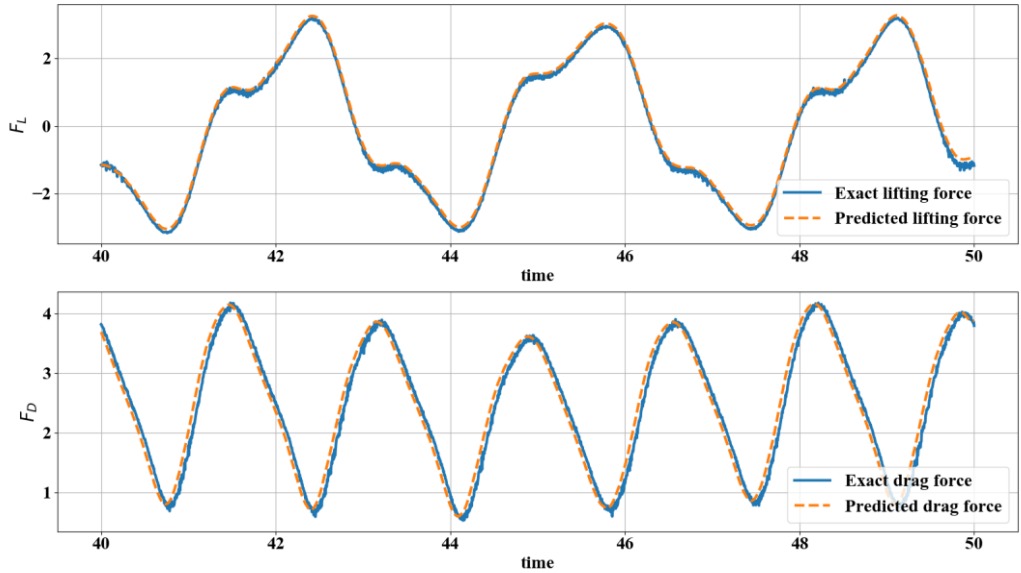

Fig. 8. PINN method for calculating the lift force and drag force

*5.4 PINN for solving wake-induced vibration of the cylinder behind a cylinder*

In this part, the wake-induced vibration of a 2-DOF cylinder which is behind a 2-DOF cylinder is investigated by PINN method. The flow field between two cylinders and surfaces of cylinders become more complex and this problem can be utilized to validate the applicability of PINN in more complicated VIV setup.

The boundary conditions are same as above and the distance between two cylinders is 6D which guarantees that the wake field of upstream structure has sufficient development space. The Reynolds number, $Re = U_\infty D/\nu$, is 9000 in this setup and flow can be viewed as turbulent. The parameters of cylinder are the damping parameter $c = 0.5183$ and the stiffness parameter $k = 2530.113$. The concrete details can be viewed in Fig. 9.

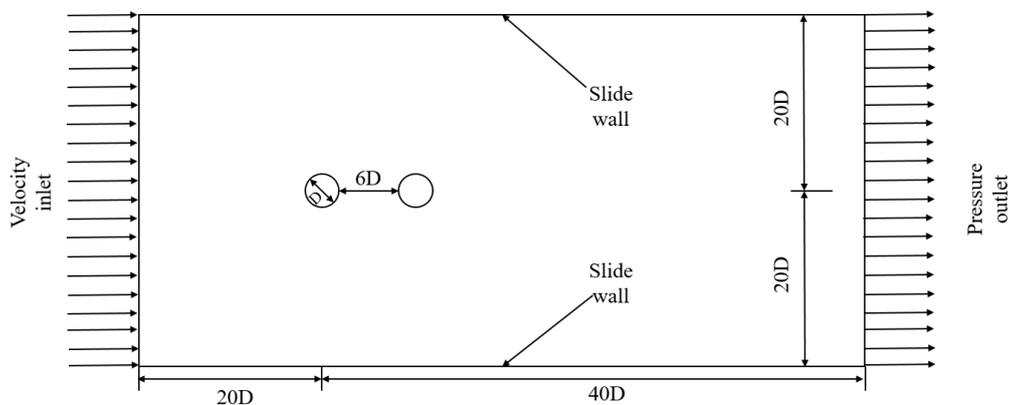

Fig. 9. The sketch of the wake-induced vibration of cylinder behind cylinder

In this section, two freedom degrees of displacements of two cylinders are

incorporated into the loss function and total loss function can be shown as:

$$e_1 = \frac{\partial u}{\partial t} + u\frac{\partial u}{\partial x} + v\frac{\partial u}{\partial y} + \frac{\partial p}{\partial x} - (v+v_t)\left(\frac{\partial^2 u}{\partial x^2} + \frac{\partial^2 u}{\partial y^2}\right) + \varsigma_{1tt} + \varsigma_{2tt}$$

$$e_2 = \frac{\partial v}{\partial t} + u\frac{\partial v}{\partial x} + v\frac{\partial v}{\partial y} + \frac{\partial p}{\partial y} - (v+v_t)\left(\frac{\partial^2 v}{\partial x^2} + \frac{\partial^2 v}{\partial y^2}\right) + \eta_{1tt} + \eta_{2tt} \quad (28)$$

$$e_3 = \frac{\partial u}{\partial x} + \frac{\partial v}{\partial y}$$

$$\begin{aligned}L_{sum} &= L_u + L_v + L_\eta + L_\varsigma + L_e \\ &= \left\|u(t^n, x^n, y^n) - u^n\right\| + \left\|v(t^n, x^n, y^n) - v^n\right\| \\ &+ \left\|\eta_1(t^n) - \eta_1^n\right\| + \left\|\varsigma_1(t^n) - \varsigma_1^n\right\| + \left\|\eta_2(t^n) - \eta_2^n\right\| + \left\|\varsigma_2(t^n) - \varsigma_2^n\right\| \\ &+ \sum_{i=1}^{3}\left\|e_i\right\|\end{aligned} \quad (29)$$

More training samples are selected in calculation region between two cylinders in order to enhance the predictive performance of PINN in WIV setup. The number of total training dataset is 45000. The entire flow fields at $t = 80s, 90s, 100s$ are adopted to validate and the results can be demonstrated in Fig. 10. It is obvious that PINN well infers the whole flow field (velocity field and pressure field) at different times from scattered velocity information. The mean square errors of velocity and pressure can be viewed in Table 2.

Table 2. Mean square errors of entire flow fields of wake-induced vibration

|  | Mean square errors | | |
| --- | --- | --- | --- |
|  | *u* | *v* | *p* |
| $t = 80s$ | $4.51\times10^{-3}$ | $2.41\times10^{-3}$ | $2.1\times10^{0}$ |
| $t = 90s$ | $3.39\times10^{-3}$ | $1.86\times10^{-3}$ | $3.47\times10^{0}$ |
| $t = 100s$ | $3.84\times10^{-3}$ | $3.32\times10^{-3}$ | $2.85\times10^{0}$ |

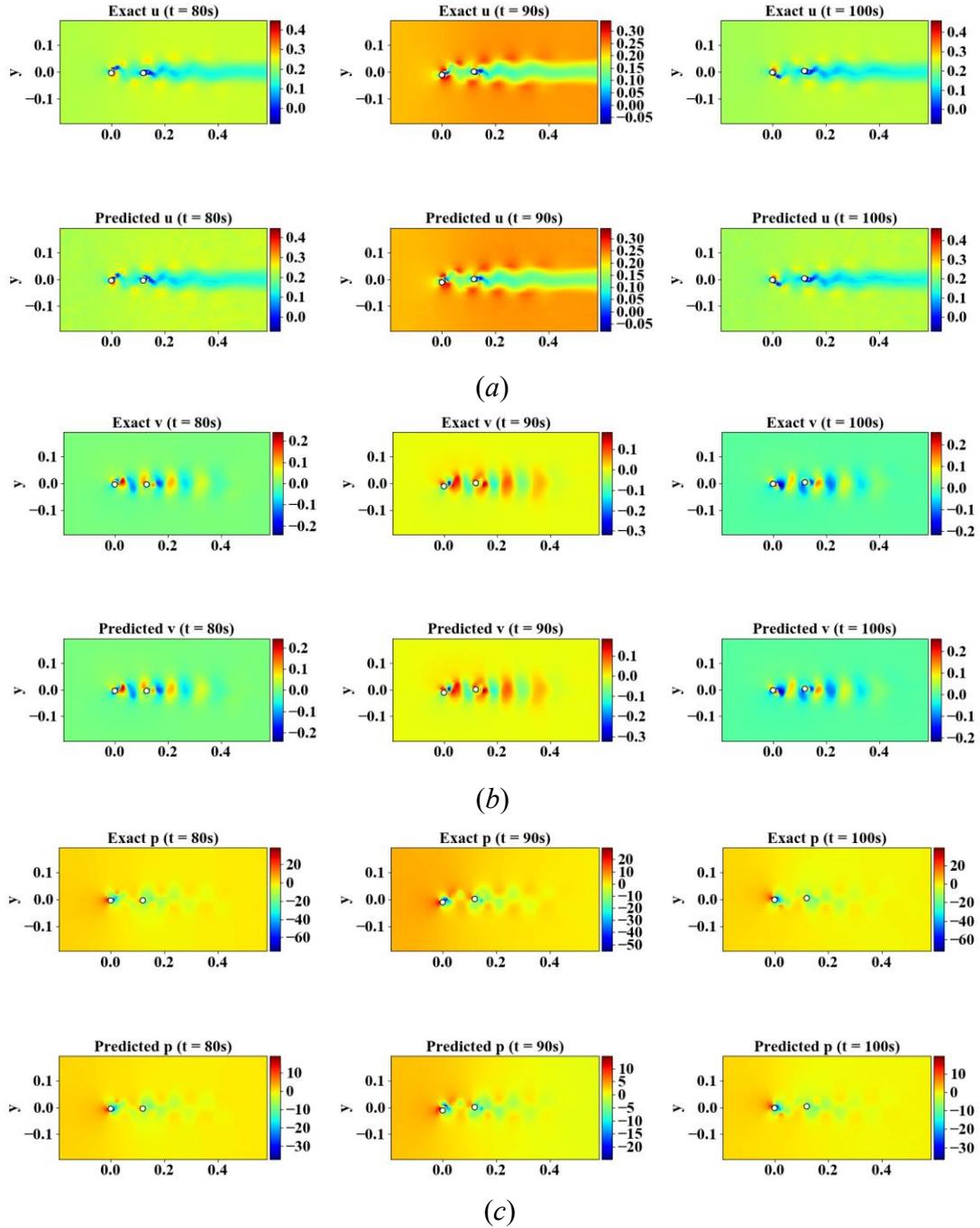

Fig. 10. PINN method for reconstructing the flow field with different times ((a) *u*, (b) *v*, (c) *p*)

With the cases investigated, the turbulent eddy viscosity is also introduced as an unknown parameter that need to be inferred. Fig. 11 indicates the inferred $v_t$ predicted by PINN and the reference $v_t$ simulated by CFD. The results show that the PINN has an effective adaptivity to approximate the unknown parameter from turbulence flow and the magnitude of mean square error at different times reaches to $10^{-4}$. This treatment represents that PINN technique could have a transformative effect for modelling turbulence closure.

Furthermore, drag force and lifting force on two cylinders predicted by PINN are also considered in this case and can be shown in Fig. 12. The mean square of errors of $F_L$ (cylinder 1), $F_D$ (cylinder 1), $F_L$ (cylinder 2), $F_D$ (cylinder 2), are $2.87 \times 10^{-4}$, $1.32 \times 10^{-4}$, $5.89 \times 10^{-5}$, $3.34 \times 10^{-5}$, respectively.

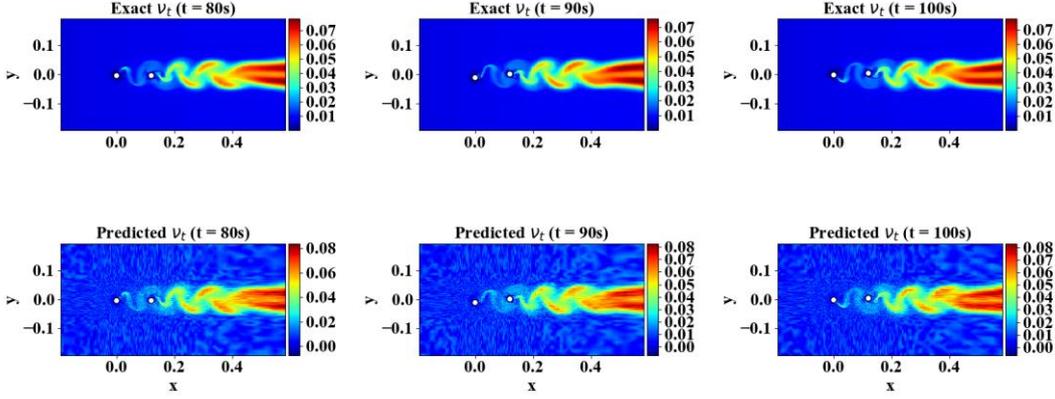

Fig. 11. PINN method for inferring eddy viscosity

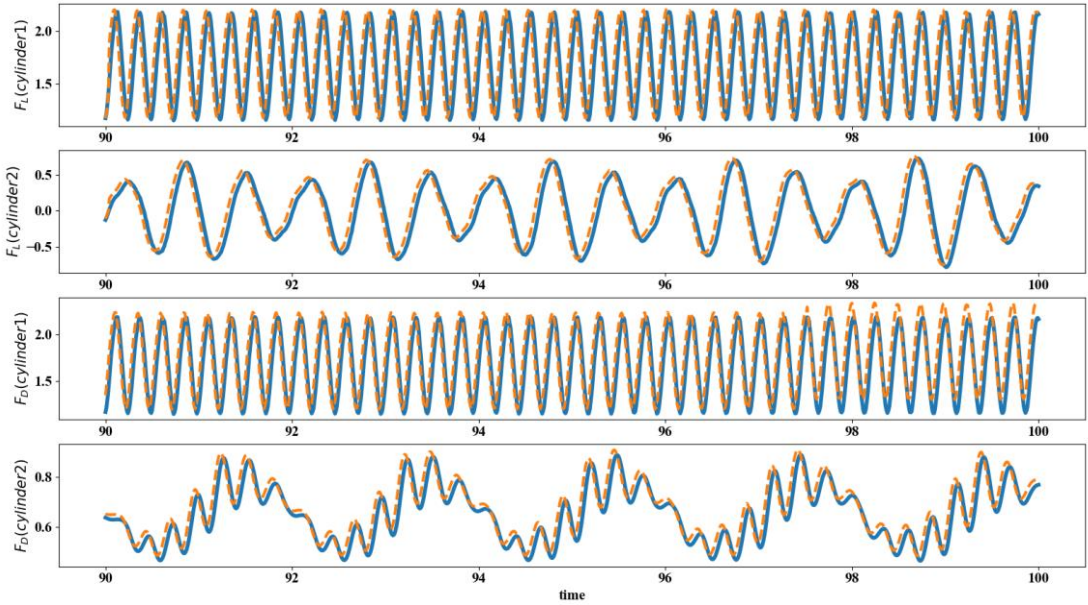

Fig. 12. PINN method for inferring lifting forces and drag forces of two cylinders

*5.5 PINN for solving unknown displacements and force based on recurrent neural network*

In this section, the displacements of cylinders are inferred directly based on lifting forces and drag forces through physics informed LSTM. The damping coefficient and stiffness coefficient are assumed as known parameters in this case. Compared to the traditional LSTM, the initial/boundary conditions (Eq. 8) and 4-th Runge-Kutta integrations (Eq. 9 and Eq. 10) are embedded into the LSTM cell which can be viewed in Fig. 13 and the comparison between inferred trajectories of two cylinders and exact ones is described in Fig. 14. Blue lines represent the exact values while red lines represent the predicted values. For the upstream cylinder, the trajectory like butterfly shape can be well inferred by the PINN. A more intriguing result is that the trajectory of downstream cylinder is more irregular, due to the complexity flow field between two cylinders, and can also well predicted by the PINN technique. The mean square errors of two trajectories can be listed in Table 3.

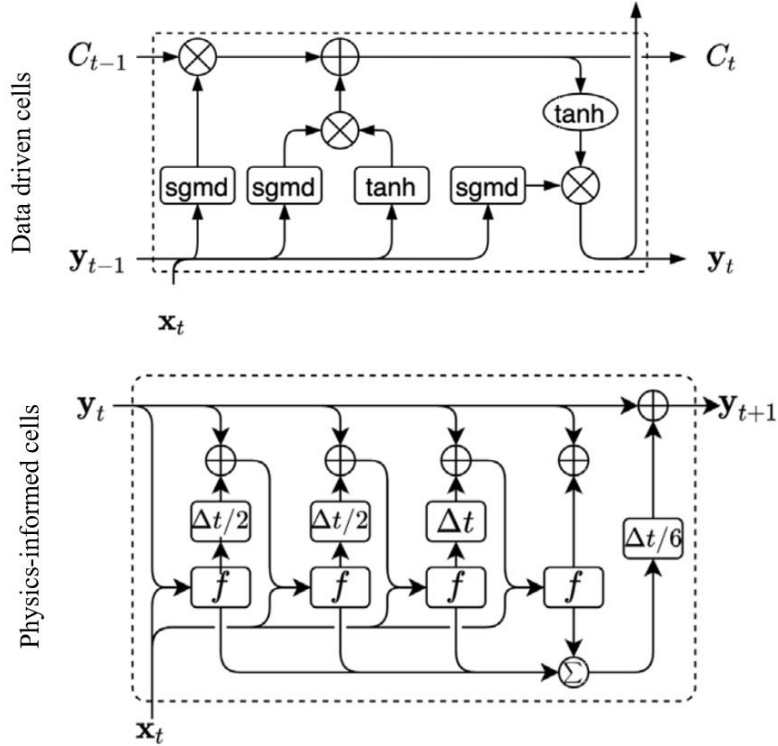

Fig. 13. Physics informed LSTM

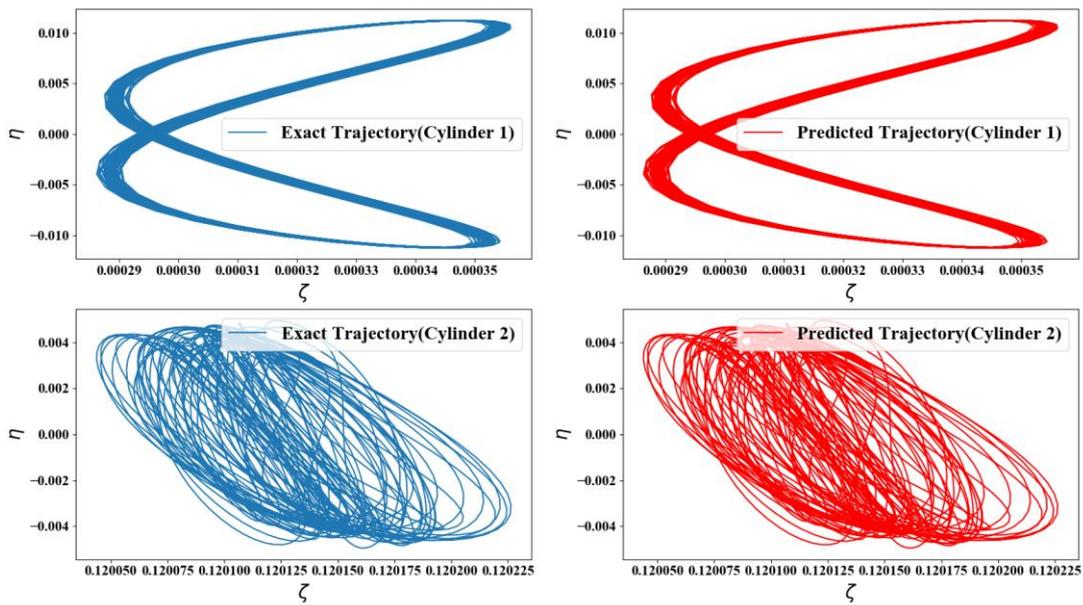

Fig. 14. PINN method for inferring trajectories of two cylinders

Table 3 The mean square error of displacements of two cylinders

|  | $\eta_1$ | $\varsigma_1$ | $\eta_2$ | $\varsigma_2$ |
|---|---|---|---|---|
| MSE | $2.1\times10^{-6}$ | $2.6\times10^{-6}$ | $1.1\times10^{-6}$ | $2.8\times10^{-6}$ |

## 6. Conclusion

In this paper, PINN based on FCNN and LSTM is adopted to solve the 2 DOF vortex-induced vibration and wake-induced vibration of cylinders under the flow in high Reynolds number or even turbulence flow. the Reynolds Average Navier-Stokes (RANS) equation, by implanting an additional turbulent eddy viscosity, coupled with structure's dynamic motion equation are also embedded into the loss function. The training samples are obtained by CFD technique. The main conclusions can be summarized as follow:

(1) PINN technique can well infer the unknown parameters (stiffness and damping coefficient) of dynamic motion equation of cylinder based on a very limited amount of training data, including force samples and displacement samples ($N$=120). The error percentages of these parameters are 0.62% and 0.02%, respectively;

(2) PINN technique can well reconstruct the whole flow field at different times including velocity field and pressure field only from scattered velocity information and the pressure information is absent. The mean square errors of flow fields reach to $10^{-3}$. Furthermore, the lift force and drag force on the cylinder can be calculated by trapezoidal law based on pressure and velocity gradients. The mean square errors of lifting force and drag force are $0.17\times10^{-4}$ and $0.63\times10^{-4}$, respectively;

(3) PINN technique has a strong applicability for solving more complicated VIV problem, called wake-induced vibration (WIV) of cylinder behind cylinder. The whole flow field, lift forces and drag forces on two cylinders can well inferred by PINN;

(4) The turbulent eddy viscosity, as an important value in turbulence, is also introduced as an unknown parameter that need to be inferred and the results show that PINN has an effective adaptivity to obtain the $v_t$, which means PINN technique could have a transformative effect for modelling the turbulence closure;

(5) The physics informed LSTM is utilized to infer the trajectories of cylinders directly based on forces. The initial/boundary conditions and 4-th Runge-Kutta integrations are embedded into the LSTM cell. The results demonstrate that the trajectories of two cylinders can be well predicted only by the force dataset.

**Acknowledgement**

This work was supported by the Fundamental Research Fund for the Central Universities of China, and the Postgraduate Research (B200203073) and Practice